\documentclass[runningheads]{llncs}
\usepackage[T1]{fontenc}
\usepackage{graphicx}
\usepackage{cite}
\usepackage{amsmath,amssymb,amsfonts}
\usepackage{algorithmic}
\usepackage{graphicx}
\usepackage{textcomp}
\usepackage{tikz}
\usepackage{amsmath}
\usepackage{multirow}
\usepackage{hyperref}
\usepackage{filecontents}
\usepackage{array}
\usepackage{hhline}
\usepackage{threeparttable}
\usepackage{adjustbox}

\usepackage{tcolorbox}

\usepackage{makecell}
\hypersetup{
    colorlinks=false,   
}

\def\BibTeX{{\rm B\kern-.05em{\sc i\kern-.025em b}\kern-.08em
    T\kern-.1667em\lower.7ex\hbox{E}\kern-.125emX}}

\usepackage{listings}
\usepackage[dvipsnames]{xcolor}
\usepackage{enumitem}

\usepackage{framed}
\definecolor{softbluegray}{rgb}{0.90,0.94,0.96}

\newenvironment{prompt}[1]{%
  \color{black}%
  \MakeFramed{\advance\hsize-\width \FrameRestore}%
  {\noindent#1}%
}{%
  \endMakeFramed
}

\begin{document}
\title{Why Not Fix It Once and for All? An Empirical Study of Multiple Patches for Vulnerability Fixes in Open-Source Software}
\titlerunning{An Empirical Study of Multiple Patches for Vulnerability Fixes}
%

\author{Weiliang Qi \and
Youpeng Li \and
Xinda Wang}
\authorrunning{W. Qi et al.}
%
\institute{University of Texas at Dallas\\
\email{\{weiliang.qi,youpeng.li,xinda.wang\}@utdallas.edu}}

\maketitle              

\begin{abstract}
Security patches for open-source software constitute a foundational resource for vulnerability remediation research and practice. However, analyzing and applying multiple patches remains challenging, especially when trying to determine at what point in a patch sequence a vulnerability is fully remediated. 
This paper presents a systematic analysis of multi-patch vulnerability fixes, focusing on their root causes, characteristics, and methods for verifying remediation status throughout the fixing process. Through a manual examination of 1,646 multi-patch fix records, we develop a taxonomy with three primary categories and six subcategories based on their causes. We then compare the distinctive characteristics of multi-patch fixes with those of single-patch fixes and analyze feature variations across categories. In addition, we assess representative vulnerability detection methods for validating complete remediation during multi-patch fixing. Our findings provide new insights into multi-patch fixes and lay a foundation for future research in this field.

\keywords{Security Patch \and{Open Source Software} \and{Vulnerability Fix}}
\end{abstract}

\section{Introduction}

\vspace{-0.3em}

To mitigate security risks in open-source software (OSS), vulnerability databases collect and publish vulnerability analyses together with the corresponding security patches. These patches serve as essential resources for both developers seeking to remediate vulnerabilities and researchers conducting vulnerability analysis. Ideally, a vulnerability is fully resolved by a single patch, leaving the software secure once the patch is applied~\cite{CNA3}. In practice, however, vulnerabilities are often addressed through multiple patches (multi-patch), which complicates the remediation process and may produce misleading outcomes~\cite{li2017large, ding2024vulnerability}.

In such cases, remediation proceeds incrementally, with different patches addressing the root cause, edge cases, or newly discovered side effects. As a result, intermediate versions may represent only partial fixes, making it difficult to determine when the vulnerability has been fully resolved. This ambiguity can lead to inaccurate assessments, noisy datasets, and flawed downstream decisions. For example, vulnerability datasets are often constructed by reversing security patches to recover vulnerable code~\cite{fan2020ac, ding2024vulnerability, bhandari2021cvefixes}. When a fix consists of multiple patches, however, reversing a later patch may yield an intermediate state rather than the original vulnerable version. Such code can exhibit both pre-patch and post-patch characteristics, making its security status difficult to determine. Naively labeling it as vulnerable or secure may therefore distort the dataset. Therefore, a systematic understanding of multi-patch scenarios is essential for improving vulnerability remediation and analysis in practice.

Although patch analysis has received substantial attention in the security community, multi-patch fixing remains underexplored. Some works~\cite{xu2022tracking, tan2022understanding} examine security patches in general, summarizing their characteristics and detection methods, while others~\cite{wang2023graphspd, wang2021patchrnn, zhou2021spi} focus on identifying and extracting security patches, such as detecting overlooked patch behavior in silent fixes. While a few studies acknowledge the existence of multi-patch fixes, they either do not provide detailed analysis~\cite{li2017large} or restrict their investigation to specific application domains~\cite{kim2023patchverif, wu2023mitigating, woo2025large}.

In this paper, we conduct a systematic study of multi-patch fixes in OSS to provide insights into this problem. Based on vulnerability patches recorded in National Vulnerability Database (NVD), we manually analyze 1,646 multi-patch fixes affecting open-source software between 1999 and 2025. We first summarize their root causes and key characteristics, and then examine the feasibility of using existing vulnerability detection techniques to determine whether the initial patch fully remediates the vulnerability. Our study is organized around following three research questions (RQs):

\noindent\textbf{RQ1: Why are multiple patches needed to fix software vulnerabilities?} Although repository management best practices, such as those in Git, recommend resolving each issue with a single commit~\cite{Git, SurveyBestPractice}, many vulnerabilities are still fixed through multiple patches.
To better understand this phenomenon, we manually analyze OSS multi-patch vulnerability samples and classify them into three main categories: multi-location fixes, fix-and-surrounding work, and defective fixes. We find that the main causes are the complexity of vulnerability contexts and remediation processes, such as vulnerabilities affecting multiple code locations or branches, which often lead developers to apply several patches. Another common cause is that the initial patch is incomplete or defective, requiring follow-up patches to complete the remediation or fix newly introduced issues. This taxonomy provides a foundation for understanding multi-patch fixing and may help developers choose more appropriate responses in practice (\S \ref{sec:categorization}).
Our manual analysis also reveals several recurring factors associated with multi-patch fixes (\S \ref{sec:cause}), including gaps between Common Vulnerabilities and Exposures (CVE) reporting rules and practice, Git workflow conventions, limited guidance on reporting security patches, pressure for rapid incident response, excessive reliance on senior committers, and vendor neglect.

\noindent\textbf{RQ2: What are the characteristics of multi-patch fixes?}
We compare multi-patch fixes with single-patch fixes in terms of their characteristics (\S \ref{sec:characteristics}). We find that multi-patch fixes are concentrated in widely used programming languages, are more prevalent in certain projects, and show a more pronounced upward trend over time. We further explore the distinctive features of the three multi-patch types. Notably, 31.7\% of multi-patch fixes involve intervals of more than one day between consecutive patches. In addition, we analyze differences in code similarity and complexity across these types. 

\vspace{0.2em}
\noindent\textbf{RQ3: How well do existing methods handle patch management in multi-patch contexts?} Given the large number of multi-patch fixes in OSS 
and the time gaps between patches that may create opportunities for attackers to conduct one-day exploitation, it is important to determine whether an initially released patch is sufficient to fully remediate a vulnerability or will require follow-up patches. To this end, we evaluate whether seven representative vulnerability detection techniques can predict whether the first patch will later evolve into a multi-patch fix (\S \ref{sec:management}), focusing on two common causes: incomplete fixes and multi-location fixes. Our results reveal the practical need for, and provide insights into, developing more effective methods to identify and manage multi-patch fixes.

In summary, this paper makes the following contributions:
\begin{itemize}[leftmargin=*]
\vspace{-0.6em}
    \item We construct and open-source\footnote{The dataset is available at \url{https://huggingface.co/datasets/XSec-Lab/MultiPatch}} a manually labeled dataset  of multi-patch vulnerability fixes 
    to support future research.
    \item We conduct the first systematic empirical study of multi-patch fixes and analyze their main causes through representative case studies.
    \item We investigate the characteristics of multi-patch fixes compared with single-patch cases, and provide insights from the perspectives of temporal patterns, project and language distributions, and patch-level properties.
    \item We assess existing methods for verifying remediation status in multi-patch settings and identify directions for future research.
\end{itemize}

\section{Background}

\textbf{Open-Source Software Patches.} 
In OSS-hosting systems such as Git, a software patch (i.e., a commit identified by a commit ID) is a change submitted to fix, update, or improve the source code of a project.
These patches can be categorized into two main types: (1) security patches, which address security vulnerabilities and are intended to enhance system security; and (2) non-security patches, which fix bugs, functional issues, and compatibility problems unrelated to security, as well as add or optimize features.
\noindent\textbf{Public Vulnerability Records.} CVE~\cite{CVE} is the most widely used public catalog of known security vulnerabilities, with each vulnerability assigned a unique CVE ID. The NVD~\cite{nvd} extends CVE records with additional information like severity scores, impact assessments, and fix-related data. Notably, security patches linked to CVEs are explicitly labeled as “patch” on NVD pages. Although a patch is typically expected to resolve a single vulnerability, we observe that many NVD entries reference multiple security patches for the same CVE. Given that CVE/NVD are widely used by cybersecurity professionals and researchers to identify, track, and analyze vulnerabilities, we use them as the primary data sources for this study.

\begin{figure}[htbp] 
    \centering 
    \vspace{-2em}
\includegraphics[width=0.9\textwidth]{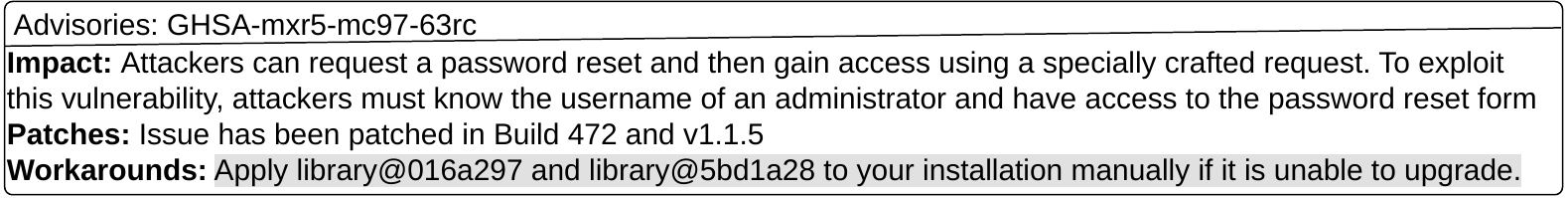} 
    \vspace{-1.2em}
    \caption{The vendor’s advisory on applying security patches for CVE-2021-32648} 
    \vspace{-2.em}
    \label{fig:patchsugg} 
\end{figure}

\noindent\textbf{Security Patch Management.}
OSS security patches linked to CVE records support vulnerability remediation in two main ways. First, patches are typically provided as Git commit URLs, enabling users to identify the exact fixed version from the commit ID and update their local code accordingly. Second, for users unable to perform a full upgrade, the relevant security changes can be applied directly using \texttt{git} \texttt{apply} or \texttt{git} \texttt{cherry-pick}. For example, as shown in \autoref{fig:patchsugg}, an OSS vendor recommends that users who cannot upgrade directly apply specific security patches instead. While applying a single patch is usually straightforward, multi-patch fixes for a single vulnerability introduce additional challenges, as discussed later.

\vspace{-1em}

\section{Data Collection}

\vspace{-0.5em}

To understand why multiple patches are needed to address a single vulnerability and how this affects practical vulnerability management, we first construct a large-scale dataset of multi-patch vulnerabilities based on CVE records.

We use the NVD JSON feeds API to collect all published CVEs as of May 2025. We then focus on CVE entries that contain reference links explicitly labeled as ``patch.'' Because not all referenced patches correspond to OSS projects with accessible source code, our study considers only OSS security patches whose URLs contain the keywords \texttt{git} and \texttt{commit}. Since OSS projects may be hosted and maintained across multiple version control platforms (e.g., GitHub and GitLab), different patch URLs may in fact refer to the same patch. Therefore, to distinguish multi-patch cases from single-patch cases accurately, we do not simply count the number of patch URLs associated with each CVE. Instead, we parse the commit IDs embedded in these URLs and use them to identify and de-duplicate patches.
Finally, retrieving patches from the referenced links poses two challenges. First, because our records span 26 years, some early patches are no longer hosted at their original repository locations, making the original links obsolete. Second, different Git hosting services use distinct APIs, which complicates large-scale data collection.
To address these issues, we manually map outdated repository addresses to their current locations and develop scripts for common platforms such as GitHub, cgit, and GitLab. Specifically, we use the GitHub API to retrieve patch metadata and code changes, while for other platforms we download and parse raw patch files.

In total, we identify 25,113 CVEs associated with OSS security patches, of which 1,646 are linked to multiple patches in the NVD, with an average of 2.55 patches per CVE.

\vspace{-1em}
\section{Categorization of Multi-Patch Fixes}
\label{sec:categorization}
\vspace{-0.5em}


This section presents a systematic manual analysis of multi-patch fixes and introduces a corresponding categorization.
The analysis was conducted by three security researchers, each with more than five years of experience in software security. First, they independently examined 100 samples randomly selected from our collected dataset and developed a preliminary taxonomy. They then discussed and refined it into a final codebook. Using this codebook, each researcher independently categorized all samples, and disagreements were resolved through discussion until consensus was reached.
In this way, we identify three main categories and six subcategories of multi-patch fixes. \autoref{tab:rq1} provides an overview of the categories and their distribution. Note that one multi-patch CVE record may belong to more than one category. Also, beyond these categories, we find 127 CVE records that include both vulnerability-introducing and vulnerability-fixing commits. Because the NVD does not distinguish vulnerability-introducing commits with a separate label, both types may be marked as patches or left unlabeled. This ambiguity can confuse users applying patches and complicate security patch collection for researchers~\cite{huntr, croft2023data, li2017large}.


\vspace{-1.5em}
\begin{table}[]
\footnotesize
\centering
\caption{Multi-patch categories and numbers of  CVEs records as of May 2025.}
\vspace{-0.8em}
\label{tab:rq1}
\setlength{\arrayrulewidth}{0.5pt}
\scalebox{0.85}{%
\begin{tabular}{l|l|c}
\noalign{\hrule height 1pt}
Category                              & Subcategory                        & \# \\ \hline 
\multirow{2}{*}{A. Multi-location fixes} & A1. Different branches or projects     & 830     \\ \cline{2-3} 
                                      & A2. Different locations within same branch & 30      \\ \hline
\multirow{2}{*}{B. Fix and related changes}    & B1. Workaround and formal fix        & 16      \\ \cline{2-3} 
                                      & B2. Fix and documentation updates 
                                & 128     \\ \hline
\multirow{2}{*}{C. Defective fixes}      & C1. Incomplete fixes                   & 641     \\ \cline{2-3} 
                                      & C2. Bug-introducing fixes           & 119     \\ 
\noalign{\hrule height 1pt}
\end{tabular}
}
\end{table}
\vspace{-3em}

\subsection{Category A - Multi-Location Fixes}
\vspace{-0.5em}

Applying multiple patches to different locations, with each patch addressing one location, is the most common multi-patch scenario. 
It typically occurs when similar vulnerable code appears in multiple methods, branches, or even separate projects. Based on the scope of the affected locations, we further divide this category into two subcategories.

\noindent\textbf{A1. Fixes for Multiple Locations Across Different Branches or Projects.}
When a vulnerability affects multiple branches or projects, the software vendor needs to address each affected instance individually. Even when similar functionality is implemented, the code context across branches may differ. In addition, Git cannot apply a single commit to multiple branches or repositories simultaneously. As a result, the vendor needs to cherry-pick or otherwise adapt the fix for every branch that requires remediation, which leads to multiple patches.

\begin{figure}[h] 
    \centering 
    \vspace{-1.5em}
    \includegraphics[width=0.8\textwidth]{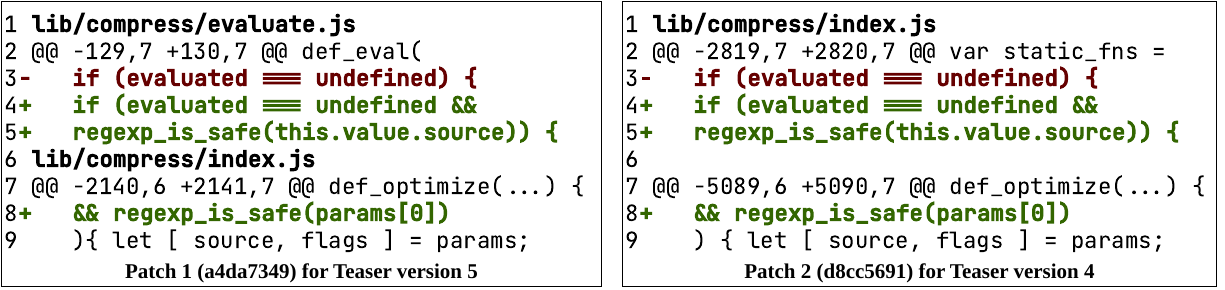} 
    \vspace{-1.25em}
    \caption{Patches for CVE-2022-25858: multi-location fixes across different branches.} 
    \label{fig:code_c1} 
\end{figure}
\vspace{-1.75em}

\noindent\autoref{fig:code_c1} shows patches for CVE-2022-25858, where separate patches were required for different branches. Teaser versions 4 and 5 are maintained in the same repository on different branches. Because a single commit cannot be applied to both branches simultaneously, the fix was first implemented in version 5 and then backported to version 4. Although the two patches follow similar logic, version 5 (left) separates compression and evaluation into two files, while version 4 keeps them in one file. Thus, the patches are similar in intent but differ in structure.

\noindent\textbf{A2. Fixes for Different Locations within the Same Branch.}
Even when a vulnerability is confined to a single branch, vendors may adopt a multi-patch strategy if the vulnerable code appears in multiple locations. This usually occurs for two reasons. First, vulnerability reports often identify only one attack path, leading vendors to patch that location first, while later analysis reveals additional vulnerable sites. Second, Git best practices~\cite{SurveyBestPractice} discourage large single commits, so vendors may split fixes into smaller commits targeting different locations. In such cases, all patches must be applied to fully remediate the vulnerability.

\begin{figure}[h] 
    \centering 
    \vspace{-2em}
    \includegraphics[width=0.8\textwidth]{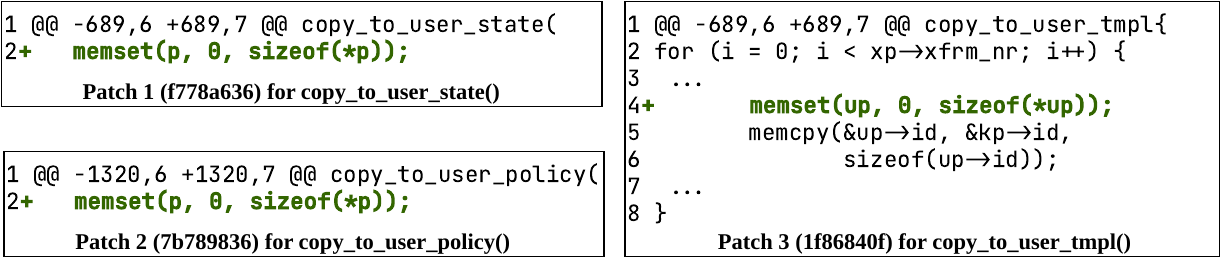} 
    \vspace{-1.25em}
    \caption{Patches for CVE-2012-6537: fixes in multiple locations within the same branch.} 
    \label{fig:code_c2} 
\end{figure}

\noindent\autoref{fig:code_c2} illustrates three commits for patching CVE-2012-6537 caused by uninitialized structures that lead to an information leak. Each patch adds explicit initialization with \texttt{memset(0)} to a different function containing similar vulnerable code, although all three functions are located in same file. While three commits use same initialization pattern, their parameters and code contexts differ.




\vspace{-2em}

\begin{prompt}
  \noindent\textbf{Insights for Applying Multi-Location Patches.} Multi-patch fixes for multi-location vulnerabilities follow the same basic rationale as single-patch fixes, but are split or ported across locations. For fixes within the same branch, all patches must be applied for complete remediation. For fixes across branches, users should apply the patch that matches their local repository configuration.
\end{prompt}

\vspace{-0.5em}
\subsection{Category B - Fix and Related Changes}
\label{sec:c2}
\vspace{-0.5em}

Some vulnerabilities require complex remediation and therefore involve progressive solutions implemented through multiple patches. In urgent cases, vendors may first release a workaround to reduce immediate risk before providing a complete fix. In large projects, development practices may also require changelogs, test cases, or other documentation to be maintained separately, leading to additional commits beyond the security fix itself.

\noindent\textbf{B1. Workaround and Formal Fix.}
Multi-patch fixes may consist of an initial workaround followed by a formal fix. The workaround serves as a temporary measure to quickly reduce exploitation risk before a complete fix is ready, while formal fix addresses root cause thoroughly. This strategy is because a full fix may take days or months, whereas immediate containment is often critical after disclosure. Workarounds also tend to have less impact on business logic, reducing risk of introducing new issues. Although both help mitigate risk, the permanent fix is generally more reliable and should be regarded as the final solution.

\begin{figure}[h] 
    \centering 
    \vspace{-2em}
    \includegraphics[width=0.8\textwidth]{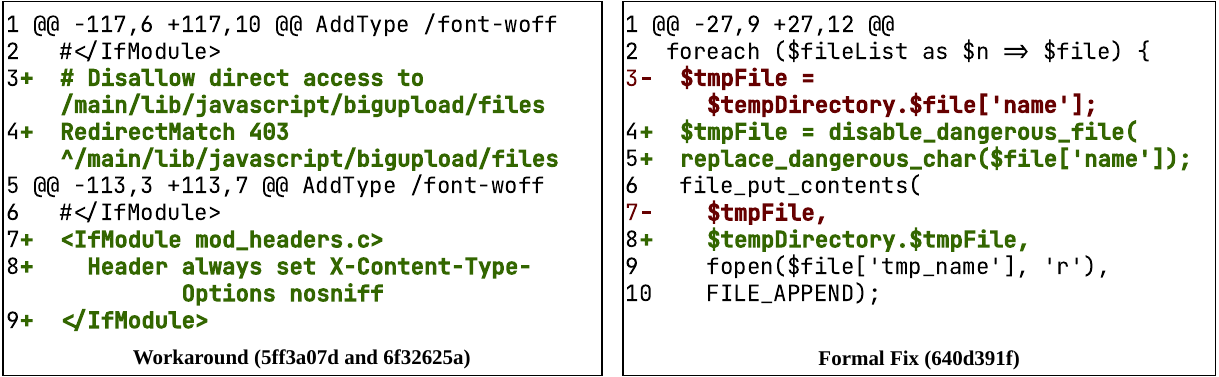} 
    \vspace{-1em}
    \caption{Patches for CVE-2023-4226: workaround and formal fix.} 
    \label{fig:code_c3} 
\end{figure}

\noindent\autoref{fig:code_c3} shows a workaround followed by a formal fix for CVE-2023-4226, an arbitrary file upload vulnerability in Chamilo LMS that can enable command execution, shell access, and stored XSS. The workaround patch (left) mitigates immediate exploitation by restricting browser parsing of risky files and blocking direct access to upload directories. The formal fix (right), released 16 days later, uses \texttt{disable\_dangerous\_file} to filter uploads and remove dangerous files.



\vspace{-2em}

\noindent\textbf{B2. Fix and Documentation Updates.}
Project development practices may require changelogs, test cases, or other documentation to be maintained separately, which leads vendors to commit these changes independently of the security fix itself. Users may either apply all related commits or cherry-pick only the security fix commit to address the vulnerability. However, when conducting patch analysis, it is necessary to carefully distinguish security-relevant commits from non-security-related ones to avoid introducing noise into the results~\cite{wang2024empirical, croft2023data, nie2023understanding}.

\begin{figure}[h] 
    \centering 
    \vspace{-2em}
    \includegraphics[width=0.8\textwidth]{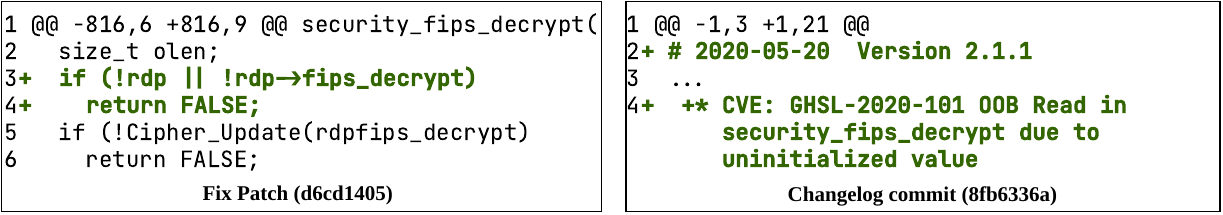} 
    \vspace{-1em}
    \caption{Patches for CVE-2020-13397: fix and documents.} 
    \label{fig:code_c4} 
\end{figure}

\noindent In \autoref{fig:code_c4}, vendor uses one patch to validate \texttt{rdp} pointer address and prevent out-of-bounds reading, while a separate commit documents this change in changelog.

     
% ...

\vspace{-2em}

\begin{prompt}
\noindent\textbf{Insights for Applying Fix and Related Change Patches.} Distinguishing related patches from actual vulnerability-fix patches is important for both practice and research. For software users, workarounds can reduce risk but may also disrupt normal functionality, such as blocking access to a vulnerable API. Once a formal fix is available, continued reliance on a workaround may therefore cause unnecessary disruption. Related patches should also be excluded from vulnerability analysis tasks. For example, in vulnerability bisection~\cite{bao2022v, zhang2025llmbisect}, documentation updates may lag behind the actual fix and obscure the true fixing point. In vulnerability code dataset construction~\cite{chen2023diversevul, bhandari2021cvefixes, fan2020ac}, documentation-related commits may be mislabeled as vulnerable samples. Excluding such related patches is therefore necessary to avoid introducing noise into the analysis.
\end{prompt}

\vspace{-0.5em}
\subsection{Category C - Defective Fixes}
\label{sec:defect}
\vspace{-0.5em}

Defects in the patch itself may also lead to multiple patches. In such cases, the initial patch may fail to fully eliminate the vulnerability or may introduce new vulnerabilities or bugs that require subsequent fixes.

\noindent\textbf{C1. Incomplete Fixes.}
\label{sec:CC1}
Ideally, vendors should implement precise vulnerability remediation, which requires a thorough understanding of both the vulnerability and the software architecture. In practice, however, the interaction between root causes and software logic often makes correct fixes difficult to develop. As a result, an initial patch may fail to fully resolve the vulnerability, leaving it to persist until a follow-up patch is released, sometimes after a substantial delay. Such incomplete remediation poses challenges for vulnerability management and software maintenance. Once vendors recognize that the initial patch is insufficient, they typically issue an additional fix, which users should apply in commit order to ensure complete remediation.

\begin{figure}[h] 
    \centering 
    \vspace{-2em}
    \includegraphics[width=0.8\textwidth]{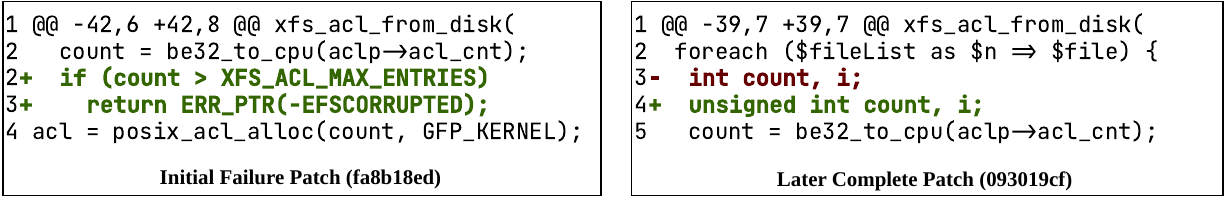} 
     \vspace{-1em}
    \caption{Patches for CVE-2012-0038: incomplete fixes.} 
    \label{fig:code_c5} 
\end{figure}

\noindent\autoref{fig:code_c5} presents CVE-2012-0038, which involves an integer overflow in \texttt{count} in the Linux XFS file system. The first patch (left), \texttt{fa8b18ed}, added a maximum-value check and exception handling. However, it overlooked that \texttt{be32\_to\_cpu()} returns a 32-bit unsigned integer, while \texttt{count} was declared as signed. As a result, \texttt{count} could be interpreted as negative in some cases and bypass the check. The vendor later released a second patch (right) that changed \texttt{count} to an unsigned integer. Notably, the complete fix was delayed by 18 days. Although the CVE had not yet been publicly disclosed, the initial patch in the open-source repository could still have exposed clues to attackers.

     

 \vspace{-2em}

\noindent\textbf{C2. Bug-Introducing Fixes.}
A defective initial patch may pose an even greater risk by introducing new bugs, including security flaws (i.e., vulnerabilities). This can create the illusion that the original issue has been fully resolved, even though the patch has introduced new problems. Vendors must then issue additional patches to remediate them.

\begin{figure}[h] 
    \centering 
     \vspace{-2em}
     \includegraphics[width=0.8\textwidth]{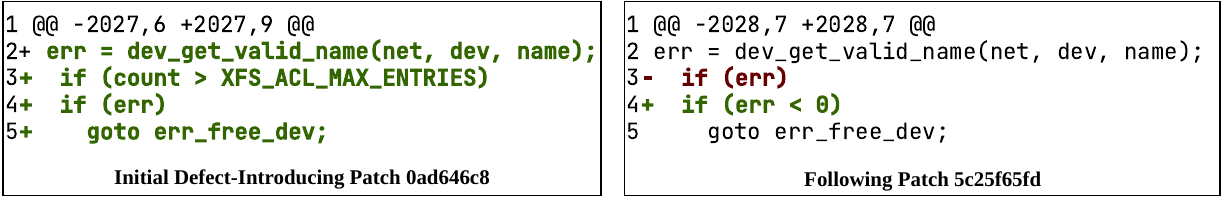} 
      \vspace{-1em}
    \caption{Patches for CVE-2018-7191: defect-introducing fixes.}
    \label{fig:code_c6} 
\end{figure}

\noindent \autoref{fig:code_c6} are patches for CVE-2018-7191 in the Linux kernel, insufficient validation of the device name before calling \texttt{register\_netd} allows a device with illegal characters to trigger a NULL pointer dereference and kernel panic, causing a denial of service. The initial patch (left) attempted to fix this by adding a device-name check. However, the developer misinterpreted the return value of \texttt{dev\_get\_valid\_name()}, assuming that only \texttt{0} indicates success, whereas any value \texttt{>=0} does. As a result, the check on \texttt{err} rejected most valid device names and introduced a new denial-of-service issue. A later patch (right) corrected the validation logic and fixed the new vulnerability.

 \vspace{-2em}

\begin{prompt}

\noindent\textbf{Insights for Handling Defective Fixes.}  For vendors, testing a security fix should verify not only that the PoC can no longer be triggered and that normal functionality is preserved, but also that the fix does not introduce new bugs or vulnerabilities~\cite{kim2023patchverif, wu2023mitigating}. OSS users should apply patches sequentially in chronological order to ensure complete remediation. For vulnerability research, defective fixes pose a distinct challenge. Prior work~\cite{ding2024vulnerability, bhandari2021cvefixes, fan2020ac, chen2023diversevul} often assumes that code after a patch is non-vulnerable and uses it as ground truth. However, code produced after a defective fix may still contain vulnerabilities, making it unsuitable as a reliable non-vulnerable sample. Therefore, although these cases are all recorded as multi-patch fixes in the NVD, they should still be distinguished by category so that appropriate measures can be taken.

\end{prompt}

\vspace{-0.25em}
\section{Contributing Factors Behind Multi-Patch}\label{sec:cause}
\vspace{-0.75em}

Based on our observations, we identify and summarize the main potential causes of multi-patch fixes as follows. 

\noindent\textbf{Gap Between CVE Rules and Practice.}
\label{sec:diffversion}
The CVE Board defines rules for assigning CVE IDs. However, ambiguous terminology in these rules may lead CVE Numbering Authorities (CNAs) to assign overly broad CVE IDs, which can in turn result in multi-patch fixes. For example, CNA Rules 3.0~\cite{CNA3} specify the following requirements for vulnerabilities affecting multiple projects:
\textit{``If multiple Products are affected by the same Independently Fixable Vulnerability, then the CNA: 
(1) MUST NOT assign more than one CVE ID if the Products are vulnerable because they share the vulnerable code. The assigned CVE ID will be shared by the vulnerable Products.
(2) SHOULD assign different CVE IDs if the Products do not share vulnerable code.
(3) SHOULD assign different CVE IDs if the CNA is uncertain whether the Products share vulnerable code.''}
However, determining whether two projects share vulnerable code is often difficult. Two projects may originate from the same third-party library or common branch, but diverge over time, making it unclear whether their vulnerable code should still be considered \textit{“shared”}. For example, CVE-2024-1394~\cite{CVE-2024-1394} affects both the \texttt{openssl} package maintained by the Golang FIPS team and Microsoft’s \texttt{go-crypto-openssl} package. Although both originate from the same OpenSSL codebase and share a memory leak, the Golang FIPS package contains an additional affected location due to platform differences. Despite these differences, the two projects were assigned the same CVE ID, and each provides its own patch, resulting in a multi-patch fix.

\noindent\textbf{Inconsistent Git Commit Granularity Practices.}
Although Git practices vary across domains, the community generally agrees that commits should be as granular as possible while remaining logically coherent~\cite{SurveyBestPractice}. Even for the same vulnerability, developers may split changes into separate commits when they affect different modules, since smaller commits are easier to manage and roll back. Because most NVD patches correspond to individual commits, this practice can naturally result in multi-patch fixes. For example, the fix for CVE-2023-35852~\cite{CVE-2023-35852} consists of two patches: one blocks abnormal absolute paths, and the other adds a configuration option for finer-grained permission control. Although such multi-patch fixes support healthier development practices, they also require clear guidance on patch application for effective security management.

\noindent\textbf{Lack of Guidance on Reporting Security Patches.}
The CVE Board sets standards for references in CVE records, such as public accessibility and long-term availability~\cite{CNA3}, but does not specify which patches should be documented. Developers may include patches unrelated to the security fix when submitting references to CNAs. For example, the record for CVE-2018-8729~\cite{CVE-2018-8729} contains one real fix and one patch that only updates the project’s README. Our statistics show that 5.8\% of multi-patch cases contain such unrelated patches, including changelogs, version updates, test cases, and README changes. These extraneous records increase the cost of patch analysis and introduce noise into automated data collection~\cite{li2017large, ding2024vulnerability, fan2020ac}.
CVE-2023-40173~\cite{CVE-2023-40173}  
record includes two patches: one creates a salting table and changes the default password, while the other updates the relevant PHP logic. 
However, they are not adjacent in the commit history. An unrecorded intermediate patch modifies the SQL file and sets the email field in the user table as a unique key. Because the patched PHP code validates the email field while the pre-fix code does not, applying only the recorded patches via \texttt{cherry-pick} may lead to logical inconsistencies.
In sum, we identify two recurring problems: CVE records may include patches unrelated to the security fix or omit prerequisite patches needed for correct application. Addressing these issues requires clearer guidance on which patches to record and how to represent patch dependencies and application context.

\noindent\textbf{Over-Reliance on Vendor Committers.}
In almost all multi-patch cases where first patch introduces a new vulnerability, the committer was already highly experienced at the time of submission, typically a founder, core developer, or early collaborator. This finding suggests that even senior contributors to large-scale projects can make mistakes, underscoring the need for rigorous code review. Notably, no CVE issued after 2020 falls into this category. This shift coincides with CVE Board’s stronger emphasis on principles for assigning new CVE IDs in CNA rules. In earlier versions~\cite{CNA2}, CVE ID assignment placed greater weight on software vendors’ opinions, which may have made vendors more inclined to assign fewer CVEs to their products, potentially due to reputation concern~\cite{wang2019detecting, li2018vuldeepecker}.

\begin{figure}[h] 
    \centering 
    \includegraphics[width=0.8\textwidth]{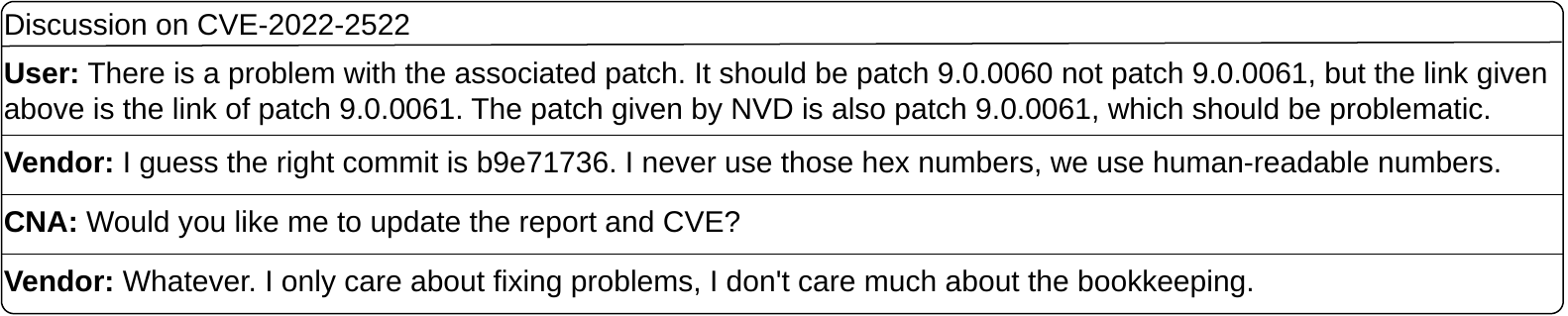} 
    \vspace{-1em}
    \caption{A discussion between OSS vendor and CNA} 
    \vspace{-2em}
    \label{fig:baddis} 
\end{figure}

\noindent\textbf{Insufficient Vendor Maintenance of Patch Records.} We find that vendors sometimes respond passively to patch maintenance, missing opportunities to correct erroneous entries. For example, in CVE-2022-2522~\cite{CVE-2022-2522}, one of the two recorded patches is actually the parent of the real security patch and is unrelated to the vulnerability fix. As shown in \autoref{fig:baddis}, a user pointed out this issue, but the vendor declined to correct it, leaving the incorrect patch in the NVD record. Although maintaining such records can be tedious, timely corrections to open-source vulnerability databases are important to the health of the community.

\vspace{-2em}

     


\section{Characteristics of Multi-Patch Fixes} \label{sec:characteristics}
\vspace{-0.5em}



This section examines the characteristics of multi-patch fixes. We first compare single-patch and multi-patch fixes across programming language, project, vulnerability type, and publication year. We then analyze the time intervals and similarities across different categories of multi-patch fixes.

\noindent\textbf{Language Distribution.}
\autoref{tab:language} presents statistics for the top-10 languages 
involved in single- and multi-patch fixes. These languages account for a larger share of multi-patch fixes, suggesting that such fixes are concentrated in widely used languages~\cite{languagerank}. 
However, this result may be influenced by data source bias. Projects written in widely used languages often receive stricter maintenance~\cite{ray2014large}, and their vendors may be more likely to split fixes into smaller commits. In addition, these languages are supported by more security analysis tools~\cite{chess2007secure}, making incomplete fixes easier to detect. Overall, these findings suggest that multi-patch fixes in open-source databases are concentrated in widely used languages, while incomplete fixes may still remain widespread in practice~\cite{woo2025large}. 

\noindent\textbf{Project Distribution.}
\autoref{tab:project} shows the distribution of single-and multi-patch fixes across the top-10 projects. Different projects exhibit clear preferences for one fixing pattern over the other. For example, Linux has 1,415 single-patch fixes but only 84 multi-patch fixes, corresponding to one multi-patch fix for every 17.8 vulnerability fixes. In contrast, ImageMagick rises from seventh place in single-patch fixes to second place in multi-patch fixes, with one multi-patch fix for every 3.2 vulnerability fixes. This pattern appears to stem primarily from its maintenance practices: ImageMagick simultaneously maintains versions 4, 5, and 6, so a single vulnerability often requires separate patches for each version. OpenSSL and CPython show similar patterns, suggesting that maintaining multiple active versions increases the likelihood of multi-patch fixes.




\vspace{-1.5em}
\begin{table}[htbp]
\caption{Top-10 programming languages and projects in single- and multi-patch fixes}
\vspace{-1.225em}
\centering
\setlength{\arrayrulewidth}{0.5pt}
\setlength{\tabcolsep}{3pt}
\begin{minipage}[t]{0.4\textwidth}
\centering
\label{tab:language}
\scalebox{0.75}{%
\begin{tabular}{c|lr|lr}
\noalign{\hrule height 1pt}
   & \multicolumn{2}{c|}{Single-Patch} & \multicolumn{2}{c}{Multi-Patch} \\ \hline
1  & C          & 29.91\% & C          & 32.02\% \\
2  & PHP        & 19.94\% & PHP        & 20.21\% \\
3  & JavaScript & 9.53\%  & JavaScript & 11.68\% \\
4  & Python     & 8.63\%  & Python     & 10.54\% \\
5  & C++        & 6.04\%  & Java       & 7.45\%  \\
6  & Java       & 4.21\%  & Go         & 5.93\%  \\
7  & Go         & 4.20\%  & C++        & 5.56\%  \\
8  & TypeScript & 3.56\%  & Ruby       & 5.05\%  \\
9  & Ruby       & 2.62\%  & TypeScript & 4.29\%  \\
10 & Rust       & 0.85\%  & Rust       & 1.90\%  \\
\noalign{\hrule height 1pt}
\end{tabular}
}
\end{minipage}
\hspace{4pt}
\begin{minipage}[t]{0.55\textwidth}
\centering
\label{tab:project}
\scalebox{0.75}{%
\begin{tabular}{c|lr|lr}
\noalign{\hrule height 1pt}
   & \multicolumn{2}{c|}{Single-patch} & \multicolumn{2}{c}{Multi-patch} \\ \hline
1  & linux          & 1415 (10.10\%) & linux          & 84 (2.08\%)  \\
2  & tensorflow     & 384 (2.74\%)   & ImageMagick    & 61 (1.51\%)  \\
3  & wireshark      & 292 (2.08\%)   & openssl        & 52 (1.28\%)  \\
4  & vim            & 190 (1.35\%)   & xwiki-platform & 33 (0.81\%)  \\
5  & gpac           & 142 (1.01\%)   & tensorflow     & 31 (0.76\%)  \\
6  & xwiki-platform & 134 (0.95\%)   & phpmyadmin     & 28 (0.69\%)  \\
7  & ImageMagick    & 133 (0.94\%)   & discourse      & 26 (0.64\%)  \\
8  & qemu           & 117 (0.83\%)   & wireshark      & 22 (0.54\%)  \\
9  & openssl        & 111 (0.79\%)   & FFmpeg         & 19 (0.47\%)  \\
10 & tcpdump        & 105 (0.74\%)   & cpython        & 18 (0.44\%)  \\
\noalign{\hrule height 1pt}
\end{tabular}
}
\end{minipage}
\end{table}
\vspace{-1em}




\noindent\textbf{Vulnerability Type Distribution.}
\autoref{tab:cwe} presents the distribution of the top-10 most common vulnerability types in multi-patch fixes. Some vulnerability types appear more frequently in multi-patch fixes. For example, CWE-200, CWE-770, and CWE-94 rank 2nd, 9th, and 10th in multi-patch fixes, but only 9th, 26th, and 16th in single-patch fixes. Compared with other high-ranking types, such as Out-of-bounds Write, these vulnerabilities often require more complex remediation. Specifically, fixing CWE-200 (Exposure of Sensitive Information to an Unauthorized Actor) usually involves both strengthening access control to prevent unauthorized access and improving sensitive data handling to block information leakage. Because these changes often affect different modules, vendors may be more likely to adopt a multi-patch approach.

\begin{table}[]
\centering
\scriptsize
\vspace{-2em}
\caption{Top-10 vulnerability types in multi-patch fixes}
\vspace{-0.8em}
\label{tab:cwe}
\setlength{\arrayrulewidth}{0.5pt}
\begin{threeparttable}
\scalebox{0.85}{%
\begin{tabular}{c|@{\hspace{0.1cm}}c@{\hspace{0.1cm}}|@{\hspace{0.1cm}}c@{\hspace{0.1cm}}|@{\hspace{0.1cm}}c@{\hspace{0.2cm}}}
\noalign{\hrule height 1pt}

CWE & Description                          & Count   & Rank (Compared w/ single) \\ \hline
79  & Cross-site Scripting                   & 519   & 1 (--)       \\
200 & Sensitive Information Exposure         & 164   & 2 ($\uparrow$7)  \\
20  & Improper Input Validation              & 164   & 3 ($\downarrow$1) \\
125 & Out-of-bounds Read                     & 157   & 4 ($\downarrow$2) \\
787 & Out-of-bounds Write                    & 133   & 5 ($\uparrow$1)  \\
119 & Buffer Overflow                        & 116   & 6 ($\downarrow$2) \\
22  & Path Traversal                         & 113   & 7 (--)      \\
89  & SQL Injection                          & 100   & 8 (--)       \\
770 & Allocation Without Limits              & 95    & 9 ($\uparrow$17)\\
94  & Code Injection                         & 83    & 10 ($\uparrow$6) \\
\noalign{\hrule height 1pt}
\end{tabular}
}

\end{threeparttable}
\end{table}
\vspace{-1.5em}

\begin{figure}[h] 
    \centering 
    \vspace{-2em}\includegraphics[width=0.7\textwidth]{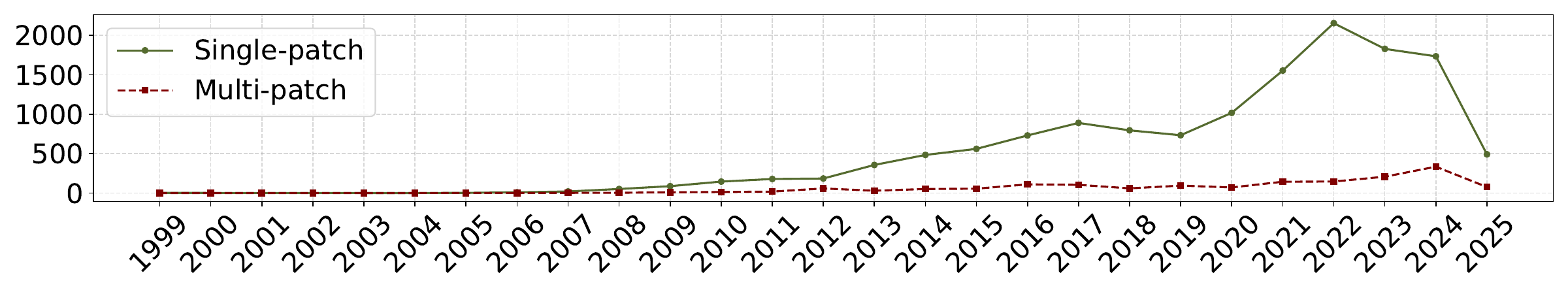} 
    \vspace{-1.25em}
    \caption{The number of single and multi-patch fixes over time} 
    \label{fig:year} 
\end{figure}

\noindent\textbf{Temporal Distribution.}
\autoref{fig:year} shows the trend of single-patch and multi-patch fixes over time. Excluding the most recent years for which some CVEs may not yet be fully disclosed, both types exhibit a clear upward trend. Notably, even in years when the number of single-patch fixes declines, multi-patch fixes continue to increase. For example, in 2023 and 2024, single-patch fixes decrease, whereas multi-patch fixes continue to grow. This pattern suggests an increasing adoption of multi-patch fixing practices.

\vspace{-2em}



\noindent\textbf{Time Interval Between Patches.}
\autoref{tab:interval} shows the time intervals for the six categories of multi-patch fixes defined in \S\ref{sec:categorization}, measured from the first patch to the last. Overall, 31.7\% of multi-patch fixes take more than one day to complete. Vendors respond more quickly when fixing vulnerabilities at multiple locations within the same branch, while porting fixes across versions or projects often causes substantial delays. This creates security risks because different versions of the same software often share similar logic: once one version is patched, attackers may search for similar flaws in unpatched versions, and delayed patch migration prolongs exposure. Delays are also common in other categories. In defective-fix cases, vendors may not realize at submission time that the initial patch is incomplete or introduces new issues, further delaying remediation.

\vspace{-1.5em}
\begin{table}[h]
\centering
\caption{Time interval of multi-patch fixes (from the first patch to the last patch). }
\scriptsize
\vspace{-0.8em}
\label{tab:interval}
\setlength{\arrayrulewidth}{0.5pt}
\scalebox{0.85}{%
\begin{tabular}{c|@{\hspace{0.1cm}}c|@{\hspace{0.2cm}}c|@{\hspace{0.2cm}}c|@{\hspace{0.2cm}}c|@{\hspace{0.2cm}}c|@{\hspace{0.2cm}}c}
\noalign{\hrule height 1pt}
Interval (days)            & \makecell{A1. Cross-\\Branch/Proj}  & \makecell{A2. In-Branch\\Multi-Loc} & \makecell{B1. Workaround\\and Fix} & \makecell{B2. Fix\\and Doc}  & \makecell{C1. \\ Incomplete}  & \makecell{C2. Bug-\\Introducing} \\ \hline
\textless~1       & 556 & 17 & 9  & 68 & 336  & 48 \\ 
1 - 30       & 137 & 6  & 5  & 29  & 145 & 43 \\ 
30 - 360     & 50  & 0  & 3  & 9   & 33  & 5  \\ 
\textgreater~360 & 10  & 0  & 0  & 1   & 4   & 1  \\ 
\noalign{\hrule height 1pt}
\end{tabular}
}
\end{table}
\vspace{-3em}

\subsubsection{Patch Similarity and Complexity.}
\begin{figure}[h] 
    \centering 
    \includegraphics[width=0.6\textwidth]{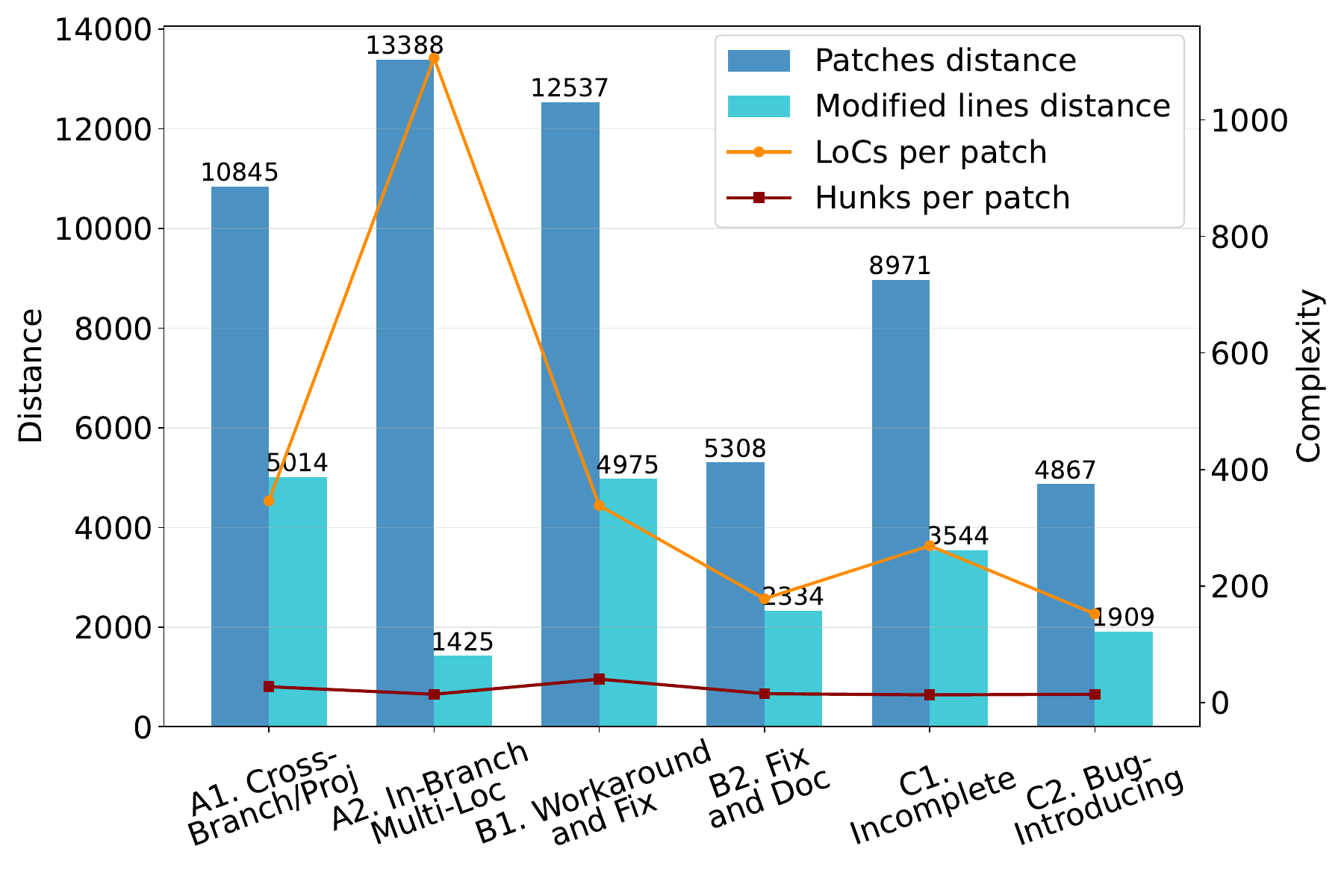} 
\vspace{-2em}
    \caption{Patch similarity and complexity across multi-patch fixes.} 

    \label{fig:simi} 
\end{figure}

We measure patch similarity using Levenshtein distance~\cite{levenshtein1966binary} at character level, which quantifies the minimum number of insertions, deletions, and substitutions required to transform one patch into another so that larger distances indicate lower similarity. The line charts further report the average number of code hunks and lines of code per patch in each category. The results in \autoref{fig:simi} reveal several non-intuitive issues. First, while fixes spanning multiple branches address the same underlying logic and might therefore be expected to exhibit high similarity, their Levenshtein distances are generally large, indicating low similarity. One possible explanation is that cross-branch porting requires additional modifications to address compatibility differences. By contrast, fixes at multiple locations within a single branch exhibit relatively high similarity, as they apply the same fix logic to similar code sites. 
We also find that bug-introducing patches and their subsequent fixes often remain highly similar, because they typically modify the same code context.

\vspace{-2em}
\vspace{-2em}


\section{Management of Multi-Patch Fixes}\label{sec:management}
\vspace{-0.5em}

Given that some vulnerabilities are not fully remediated by a single patch (\S\ref{sec:categorization}), an important practical question is whether an initial patch has completely fixed the vulnerability. As summarized in \autoref{tab:rq1}, two factors may cause multiple patches for complete remediation: i) \textit{Incomplete Patches:} the initial patch fails to fully correct the vulnerable functionality (C1); ii) \textit{Multi-Location Patches:} the initial patch fails to cover all vulnerable locations (A1 and A2). This section assesses the extent to which existing vulnerability analysis techniques can detect whether an initial patch will later require follow-up patches from these two perspectives.

\vspace{-0.5em}
\subsection{Vulnerability Detection for Incomplete Patches}
\vspace{-0.5em}

An incomplete patch leaves the code vulnerable even after the patch is applied. If an ideal vulnerability detection (VD) oracle is available, we could use it to determine whether the patched code still contains the vulnerability and thus whether the patch is complete. In this subsection, we evaluate whether existing VD approaches are capable of serving this purpose.

\noindent\textbf{Setup.}
We select six representative VD models across different architectures:
\textit{\textbf{CodeBERT}}~\cite{feng2020codebert} and \textit{\textbf{UniXcoder}}~\cite{guo2022unixcoder} are general-purpose pre-trained models for code understanding that also demonstrate strong performance on VD tasks~\cite{lu2021codexglue}. \textit{\textbf{Devign}}~\cite{zhou2019devign} and \textit{\textbf{ReVeal}}~\cite{chakraborty2021deep} are graph neural network based models designed for vulnerability detection. \textit{\textbf{LineVul}}~\cite{fu2022linevul}, \textit{\textbf{VulBERTa}}~\cite{hanif2022vulberta}, and \textit{\textbf{PDBERT}}~\cite{ding2024vulnerability} leverage pre-trained language model architectures with enhancements for vulnerability detection.
%
We exclude pre-trained large language models (LLMs) because our dataset is collected from public vulnerability databases, making it difficult to rule out data leakage from pre-training corpora.
These models are trained following standard procedures in prior VD studies and used to predict whether a patch completely fixes a vulnerability. Specifically, \textit{to detect incomplete fixes, a well-performing model should classify the code before the final fixing patch as vulnerable and the code after it as non-vulnerable}. 

\noindent\textbf{Dataset.} 
We conduct the test on our collected data from category (C1) as incomplete patches of multi-patch fixes. The code before the final fix is labeled as vulnerable, and the code after the final fix as non-vulnerable. Since our goal is to evaluate whether existing VD models can detect vulnerabilities that remain after an incomplete fix, we train these models following the standard data construction routine used in prior work, such as Big-Vul~\cite{fan2020ac}. Specifically, we collect vulnerable and non-vulnerable samples from single-patch cases in the NVD and split them into training, validation, and test sets in 8:1:1, as shown in \autoref{tab:incomp}.

\vspace{-1em}
\begin{table}[ht] \centering
\caption{Composition of incomplete-patch detection dataset.}
\footnotesize
\vspace{-0.8em}
\label{tab:incomp}
\setlength{\arrayrulewidth}{0.5pt}
\scalebox{0.85}{%
\begin{tabular}{@{\hspace{0.2cm}}c|@{\hspace{0.5cm}}c@{\hspace{0.5cm}}c@{\hspace{0.5cm}}c|@{\hspace{0.5cm}}c@{\hspace{0.2cm}}} 
\noalign{\hrule height 1pt}
Dataset & Train (VD)  & Valid (VD) & Test (VD)  & Test (multi)\\\hline
Vulnerable     & 8301   & 1037  & 1037        & 738        \\ 
Non-Vulnerable & 243608 & 30451 & 30451       & 556        \\ 
\noalign{\hrule height 1pt}
\end{tabular}
}
\end{table}
\vspace{-2em}

\noindent\textbf{Result.}
\autoref{tab:C1_res} reports the performance of each VD model on detecting incomplete patch fixes. Accuracy and F1 reflect overall performance. True positive rate (TPR), equivalent to recall, measures model’s ability to identify remained vulnerabilities after incomplete fixes. True negative rate (TNR) represents its ability to recognize fully fixed code as non-vulnerable. We also compare these results with performance on single-patch data to assess the extent to which existing VD models degrade when detecting vulnerabilities left by incomplete fixes.

\vspace{-1em}
\begin{table}[ht] \centering
\caption{VD models performance on detecting incomplete fixes requiring multi-patch}
\footnotesize
\label{tab:C1_res}
\vspace{-0.8em}
\setlength{\arrayrulewidth}{0.5pt}
\scalebox{0.85}{%
\begin{tabular}{l|@{\hspace{0.1cm}}c@{\hspace{0.2cm}}c@{\hspace{0.2cm}}c@{\hspace{0.2cm}}c@{\hspace{0.1cm}}|@{\hspace{0.1cm}}c@{\hspace{0.2cm}}c@{\hspace{0.2cm}}c@{\hspace{0.2cm}}c}
\noalign{\hrule height 1pt}
\multirow{2}{*}{Model} & \multicolumn{4}{@{\hspace{-0.1cm}}c|@{\hspace{0.1cm}}}{Vulnerability Detection} & \multicolumn{4}{c}{Incomplete Patch Detection} \\  
                       & Acc   & TPR     & TNR    & F1     & Acc     & TPR     & TNR     & F1    \\ \hline
CodeBERT               & 96.94 & 38.54   & 98.93  & 45.38  & 45.43($51.51\downarrow$)  & 10.16($28.38\downarrow$)  & 92.42($6.51\downarrow$)  & 17.54($27.84\downarrow$) \\
PDBERT                 & 96.94 & 38.25   & 98.94  & 45.16  & 45.28($51.66\downarrow$)  & 10.30($27.95\downarrow$)  & 91.88($7.06\downarrow$)  & 17.69($27.47\downarrow$) \\
UniXcoder              & 96.85 & 39.98   & 98.78  & 45.53  & 45.36($51.49\downarrow$)  & 13.82($26.16\downarrow$)  & 87.36($11.42\downarrow$)  & 22.42($23.11\downarrow$) \\
VulBERTa               & 96.79 & 3.37   & 96.81  & 6.48   & 42.96($53.83\downarrow$)  & 0.27($3.10\downarrow$)  & 99.82($3.01\uparrow$)  & 0.53($5.95\downarrow$) \\
Linevul                & 96.93 & 10.02  & 99.88  & 75.36  & 43.73($53.20\downarrow$)  & 2.85($7.17\downarrow$)  & 98.19($1.69\downarrow$)  & 5.46($69.90\downarrow$) \\
Devign                 & 69.57 & 73.69  & 69.42  & 13.77  & 43.56($26.01\downarrow$)  & 63.17($10.52\downarrow$)  & 33.75($35.67\downarrow$)  & 42.73($28.96\uparrow$) \\
ReVeal                 & 71.63 & 70.52  & 71.68  & 14.08  & 48.85($22.78\downarrow$)  & 50.18($20.34\downarrow$)  & 48.20($23.48\downarrow$)  & 39.54($25.46\uparrow$) \\
\noalign{\hrule height 1pt}
\end{tabular}
}
\end{table}
\vspace{-1em}

\noindent We find that the TPR of all models on incomplete-fix detection drops by 7.17\% to 28.38\% compared with vulnerability detection on single-patch cases, indicating that VD models trained on conventional data struggle to identify vulnerabilities that remain after incomplete fixes. The TNR of transformer-based models also declines consistently, suggesting that their performance is affected by the distinctive characteristics of multi-patch data. Our examination of failure cases shows that these models often assign the same label to the pre-fix, intermediate, and post-fix versions of functions involved in incomplete fixes. Because security patches typically introduce only partial changes to a function, the models often fail to distinguish among these closely related variants. Notably, all models achieve accuracy and F1 scores below 50\% on incomplete-fix detection, which is worse than random guessing. These results highlight the need for more effective methods to predict whether a patch will evolve into a multi-patch fix due to incomplete remediation. 


\vspace{-0.5em}
\subsection{Vulnerable Code Clone Detection for Multi-Location Patches}
\vspace{-0.5em}

Vulnerabilities recurring at different code locations may require multiple patches for complete remediation. The first patch may fix the vulnerability at one location, while similar code sites remain vulnerable. Vulnerable code clone (VCC) detectors are designed to identify similar code sites and determine whether they contain vulnerabilities. In this subsection, we evaluate whether such techniques can identify cases in which additional patches are needed at other locations.

\noindent\textbf{Method.}
Since our test samples are collected from the NVD, many proprietary clone detection tools may already contain fingerprints of the recorded vulnerabilities. To avoid data leakage, we select two popular open-source tools that support building vulnerability fingerprints from scratch, {\textit{\textbf{ReDebug}}}~\cite{jang2012redebug} and {\textit{\textbf{FIRE}}}~\cite{feng2024fire}.

\noindent\textbf{Setup.} We use the multi-location patch samples manually annotated as categories A1 and A2 for evaluation. Specifically, we use the first patch in each patch sequence to generate vulnerability fingerprints for the two tools, and use the remaining patches to construct the test set. This yields a dataset of 252 vulnerability signatures and 266 test samples. We then apply the two VCC tools to \textit{determine whether recurring vulnerabilities are present by classifying the other code site before each subsequent fixing patch as vulnerable and the code after the patch as non-vulnerable.}


\noindent\textbf{Result.}
\autoref{tab:A_res} reports performance of VCC methods on multi-location vulnerabilities in multi-patch fixes, alongside general VCC results reported in FIRE~\cite{feng2024fire}. Both ReDebug and FIRE show clear performance drops in the multi-location setting. Particularly, FIRE’s TPR decreases from 90 to 52.46. One possible reason is that vulnerable locations in multi-patch cases differ substantially, making recurring vulnerabilities harder to detect. The FPR of both tools also decreases, suggesting that they tend to miss related vulnerabilities rather than over-predict them. These results indicate that both tools struggle to capture recurring vulnerability signatures across different code locations in multi-patch fixes.

\vspace{-1em}
\begin{table}[ht] \centering
\caption{Vulnerable code clone detection on multi-location patches}
\vspace{-0.9em}
\footnotesize
\label{tab:A_res}
\setlength{\arrayrulewidth}{0.5pt}
\scalebox{0.85}{%
\begin{tabular}{@{\hspace{0.1cm}}l@{\hspace{0.1cm}}|@{\hspace{0.1cm}}c@{\hspace{0.2cm}}c@{\hspace{0.1cm}}|@{\hspace{0.1cm}}c@{\hspace{0.2cm}}c@{\hspace{0.1cm}}}
\noalign{\hrule height 1pt}
Model     & VCC\_TPR  & VCC\_FPR  & Multi\_TPR  & Multi\_FPR \\ \hline
ReDebug   & 37.56 & 20.05  & 16.75($\downarrow$20.81)            & 0.49($\downarrow$19.56)     \\
FIRE      & 90.00 & 8.57  & 52.46($\downarrow$37.54)            & 5.23($\downarrow$3.34)   \\ 
\noalign{\hrule height 1pt}
\end{tabular}
}
\end{table}
\vspace{-2em}

\vspace{-0.8em}
\section{Related Work}
\vspace{-0.8em}

\noindent\textbf{Patch-Based Vulnerability Analysis.}
Security patches play an important role in vulnerability analysis. Prior work~\cite{bao2022v, tang2023neural} infers vulnerability life cycles from patches, while other studies~\cite{yan2020just, pornprasit2021jitline, hoang2019deepjit, nguyen2024code} use patch information to determine whether patches introduce vulnerabilities. Existing real-world vulnerability datasets~\cite{zhou2019devign, chakraborty2021deep, fan2020ac, chen2023diversevul, ding2024vulnerability} are often constructed by reverting patches, and synthesized datasets also rely on patches to capture vulnerability patterns. 
Researchers have proposed fuzzing-based~\cite{huang2024titan, fioraldi2020afl++, klees2018evaluating, bohme2017directed}, code clone-based~\cite{kim2017vuddy, xiao2020mvp}, and deep learning-based~\cite{li2018vuldeepecker, li2021sysevr, fu2022linevul, nguyen2022regvd} methods for vulnerability detection. For example, MVP~\cite{xiao2020mvp} uses patch-signature matching, VUDDY~\cite{kim2017vuddy} uses code clone detection, and Magma~\cite{hazimeh2020magma} builds fuzzing benchmarks by reverse-integrating security patches. Therefore, a correct understanding of security patches is important for the quality of software security datasets and subsequent research.

\noindent\textbf{Patch Analysis.}
A series of efforts has been made on software patch analysis. Zhong et al.~\cite{zhong2015empirical} analyze 9K bug fixes across six Java OSS projects, while Sliwerski et al.~\cite{sliwerski2005changes} examine relationship between bug-introducing changes and bug-fixing patches in two OSS projects. Iannone et al.~\cite{iannone2022secret} investigate 3K security patches in the NVD, discussing their causes and fix processes. VFCFinder~\cite{dunlap2024vfcfinder} mines vulnerability-fixing patches from vulnerability reports. PatchRNN~\cite{wang2021patchrnn} and SPI~\cite{zhou2021spi} use deep learning to identify security patches, while GraphSPD~\cite{wang2023graphspd} leverages code property graphs over pre- and post-patch code to improve detection. Xu et al.~\cite{xu2017spain} propose a binary-level framework for identifying security patches and summarizing vulnerability patterns.
However, prior studies on multi-patch fixes either focus on general bugs or functional patches rather than security fixes, or are limited to specific domains. Li et al.~\cite{li2017large} discuss multi-patch fixes in a large-scale study of security patches, but analyze only 100 samples. Gu et al.~\cite{gu2010has} study bug-fix patches in three OSS projects. Wu et al.~\cite{wu2023mitigating} and Kim et al.~\cite{kim2023patchverif} respectively examine patch correctness in Linux and robotics systems.


\vspace{-0.8em}
\section{Discussion and Conclusion}
\vspace{-0.8em}

Our classification of multi-patch fixes relies on manual analysis, which introduces potential threats to validity. Since our analysts are not the original developers of the projects, they may misjudge the causes of certain vulnerabilities. 
To mitigate these threats, three researchers independently analyzed each sample and discussed disagreements in complex cases until reaching consensus. We also maintained a codebook throughout the process to formalize the accumulated knowledge, and any modification to the codebook required agreement from all participants.
Moreover, our analysis is based on public records provided by the NVD. Although NVD records are maintained by NIST analysts, errors may still occur given the large scale of the database. For multi-patch data, we remove anomalous items during manual annotation. Also, because some vulnerabilities are fixed silently, certain fix records may be missing~\cite{woo2025large}. We leave large-scale analysis of patches from other resources to future work. We also hope that our study motivates further advances in patch mining and detection, enabling the discovery of more silent fixes and benefiting future patch analysis research.


To the best of our knowledge, this paper presents the first systematic empirical study of multi-patch fixes. We analyze 1,646 multi-patch fixes from the NVD, manually identify their causes, and classify them into six categories. We further compare multi-patch and single-patch fixes, examine unique characteristics, and evaluate whether existing vulnerability detection and code clone-based methods can predict when an initial fix will require follow-up patches. Our results show that current techniques struggle to handle multi-patch fixes, highlighting the need for further research. Overall, our findings provide practical insights for the security community and a foundation for future work on multi-patch fixes.

\subsubsection{\ackname} This research is partially supported by the National Science Foundation (NSF) grant CNS-2450602. Any opinions, findings, and conclusions or recommendations expressed in this material are those of the author(s) and do not necessarily reflect the views of the NSF. 
%


%
%
%
%
\bibliographystyle{splncs04}
\bibliography{main}

\end{document}